\documentclass[11pt]{article}

\usepackage[T1]{fontenc}
\usepackage[utf8]{inputenc}
\usepackage[margin=1in]{geometry}
\usepackage{microtype}

\usepackage{authblk}

\usepackage{amsmath}
\usepackage{amssymb}
\usepackage{amsthm}
\usepackage{mathtools}
\usepackage[mathscr]{euscript}

\usepackage{graphicx}
\graphicspath{{images/}}

\usepackage{booktabs}
\usepackage{multirow}
\usepackage{array}
\usepackage{tabularx}
\usepackage{longtable}
\usepackage{rotating}
\usepackage{float}

\usepackage[
  font=small,
  labelfont=bf,
  tableposition=top
]{caption}

\usepackage{algorithm}
\usepackage{algpseudocode}

\usepackage{xcolor}
\usepackage{enumitem}

\usepackage[authoryear,round]{natbib}

\usepackage[
  colorlinks=true,
  citecolor=blue,
  linkcolor=blue,
  urlcolor=blue
]{hyperref}

\theoremstyle{plain}
\newtheorem{theorem}{Theorem}[section]
\newtheorem{lemma}[theorem]{Lemma}

\newtheorem{corollary}[theorem]{Corollary}

\theoremstyle{definition}

\theoremstyle{remark}

\def\av{\boldsymbol a}

\def\ev{\boldsymbol e}

\def\yv{\boldsymbol y}

\newcommand{\Sigmav}{\boldsymbol{\Sigma}}
\newcommand{\Var}{\operatorname{Var}}

\def\Abf{\mathbf A}
\def\Bbf{\mathbf B}

\def\Dbf{\mathbf D}

\def\Pbf{\mathbf P}
\def\Qbf{\mathbf Q}

\def\Xbf{\mathbf X}
\def\Ybf{\mathbf Y}

\def\1v{\mathbf 1}
\def\0v{\mathbf 0}
\def\Id{\mathbf I}

\newcommand{\argmin}{\operatornamewithlimits{argmin}}

\newcommand{\zhao}[1]{\textcolor{black}{#1}}

\definecolor{mypink}{RGB}{227,119,194}
\definecolor{myyellow}{RGB}{188,189,34}
\definecolor{mycyan}{RGB}{23,190,207}
\definecolor{myred}{RGB}{214,39,40}
\definecolor{mygreen}{RGB}{44,160,44}
\definecolor{myorange}{RGB}{255,127,14}
\definecolor{myblue}{RGB}{31,119,180}
\definecolor{mypurple}{RGB}{148,103,189}
\definecolor{mylotus}{RGB}{140,86,75}
\definecolor{mygray}{RGB}{127,127,127}

\providecommand{\keywords}[1]{%
  \par
  \vspace{0.5em}
  \noindent\textbf{Keywords: }#1
  \par
}

\title{A Consistent Feature Screening Approach for Tensor Responses
with Applications to Genome-Wide Facial Shape Association}

\author[1]{Shaofei Zhao}
\author[2]{Zuofeng Shang}
\author[3,4]{Seth M. Weinberg}
\author[5,6,7]{Peter Claes}
\author[3,4]{John R. Shaffer}
\author[1]{%
  Guifang Fu\thanks{%
    Corresponding author:
    \href{mailto:gfu@binghamton.edu}{gfu@binghamton.edu}%
  }%
}

\affil[1]{%
  Department of Mathematics and Statistics,
  Binghamton University
}

\affil[2]{%
  Department of Mathematical Sciences,
  New Jersey Institute of Technology
}

\affil[3]{%
  Center for Craniofacial and Dental Genetics,
  University of Pittsburgh
}

\affil[4]{%
  Department of Oral and Craniofacial Sciences,
  University of Pittsburgh
}

\affil[5]{%
  Department of Electrical Engineering,
  ESAT/PSI, KU Leuven
}

\affil[6]{%
  Department of Human Genetics,
  KU Leuven
}

\affil[7]{%
  Murdoch Children Research Institute
}

\date{}

\begin{document}

\maketitle

\begin{abstract}
As data collecting technologies advance, data structures are getting more and more complex, from single vectors to multi-dimensional tensors. {\color{black}This article is motivated by a variable selection problem to detect important genes from an ultrahigh dimensional pool that are associated with  human facial shape variations. We propose a data-driven trimmed feature screening method based on a tensor ridge regression model (TrimTenRidge) through setting thresholds on the tensor coefficients to perform a feature screening procedure.

Unlike existing approaches, the TrimTenRidge does not require any sparse structures. In addition, it not only detects important predictors but also locates specific regions/components of the tensor response that are associated with each of the selected predictors. We prove the theoretical selection consistency and also assess its empirical performance through various simulation settings.} The approach copes with ultra-high dimensional predictors and tensor responses simultaneously and contributes to the literature from theoretical, methodological, and five applicational aspects. We further apply the TrimTenRidge approach to genome-wide human facial shape data, from which the entire facial shapes form a $2,342\times 7,160\times 3$ tensor, and we successfully detect several novel genetic loci and also confirm some existing findings that are associated to facial shape.
\end{abstract}


\section{Introduction}

Tensor data, multidimensional or multi-way measurements are attracting more and more attention in various fields, such as imaging \citep{li2010tensor}, multi­omics \citep{bersanelli2016methods}, microbiome \citep{martino2021context}, fMRI and EEG \citep{song2017multilinear}, radar signal processing \citep{nion2010tensor}, natural language parsing \citep{collins2012tensor}, and others \citep{cao2014tensor}. The existing tensor linear regression models have been focused mainly on prediction or estimation \citep{zhou2013tensor, sun2017store, li2017parsimonious, lock2018tensor, raskutti2019convex}, while relatively little work has been done on variable selection for ultrahigh dimension settings.

The motivating example is the detection of genetic factors contributing to inter-individual variation in human facial shape through genome-wide association studies (facial shape-GWAS in abbreviation)  \citep{claes2018genome}. The facial shape for each of the 2, 342 unrelated participants of Eu­ropean ancestry was described as a mesh of 3D XYZ-­coordinate of 7,160 vertices in .obj format, and then aligned to establish homology \citep{claes2014modeling}. Since the candidate predictor pool consists of 9, 478, 608 single nucleotide polymorphisms (SNPs), good feature screening approaches that are feasible for tensor responses are needed.

Challenges come from facial shape-­GWAS data due to its nature of ‘doubled’ high dimensionalities and ‘doubled’ complex structures: On the phenotype side, facial shape is a complex, multidimensional, and polygenic trait. As noted by \citet{claes2018genome}, many shape studies have represented facial shapes either by simple measures or low dimensional vectors \citep{adhikari2016genome, paternoster2012genome, bonfante2021gwas, xiong2019novel}, which may not capture the true morphological complexities of biological shapes. On the genetic hand, genotype data has non-­polynomial dimensionality or “ultrahigh dimension,” where the number of variables (i.e., SNPs) is in exponential level of the number of observations (i.e., study participants).

\citet{fan2008high} demonstrated for a univariate response that even a simple classification using all the predictors can be as poor as random guessing due to noise accumulation for ultrahigh-dimensional settings. Several feature screening approaches have been proposed to cope with the ultrahigh dimensional data of $p>>n$ with theoretical guarantees of consistency in variable selection \citep{candes2007dantzig, meinshausen2009lasso, bickel2009simultaneous, zhang2008sparsity, fan2008sure, zhu2011model}. However, existing feature screening approaches mainly focused on univariate responses or at most multivariate responses, and there do not exist any feature screening studies in the literature that are designed for a tensor response yet.

In this paper, we propose a data-driven trimmed feature screening method based on a tensor ridge regression model (TrimTenRidge) through setting thresholds on the tensor coefficients to perform a feature screening procedure. The contribution of the TrimTenRidge model is (1) Theoretically speaking, its selection consistency is guaranteed without any sparsity assumptions; (2) Methodologically speaking, it enables the feature screening procedure for tensor responses in ultrahigh dimensional settings; and (3) Practically speaking, it brings at least five exciting breakthroughs. Firstly, the lift of sparsity assumptions enables detection of a large amount of SNPs that have small but non-zero effects. Secondly, we not only detect important SNPs/genes, but also locate specific facial regions from the outcome end point that those selected SNPs/genes are associated with. Thirdly, we jointly put all SNPs in each of the chromosomes into one model and overcome limitations of single-SNP models. Fourthly, we model the entire facial shape as a high-dimensional tensor structure where less information is lost compared to separately and individually modeling each small segment. Fifthly, we want to emphasize that one of the rare advantages of the TrimTenRidge method is that it can detect not only those genes that are associated with the entire face but also some other genes that are only associated with certain local facial regions.

While there are some existing approaches that are relevant to the tensor regression, the proposed TrimTenRidge approach is different from these. Specifically, the existing tensor regression approaches can be summarized as 1) scalar response-tensor predictor \citep{zhou2013tensor,li2018tucker}, the reverse of our focus; 2) tensor response-tensor predictor \citep{lock2018tensor,raskutti2019convex}, different from our motivating problem; and 3) tensor response-matrix predictor \citep{sun2017store, li2017parsimonious}. The third category has the same focus as this article; however, all of these approaches were established on various sparse structures, without the capability to handle the case that has a large amount of small non-zero effects like genome-wide association studies usually have. Specifically, \citet{sun2017store} assumed the canonical decomposition/parallel factors (CANDECOMP/PARAFAC or CP) factorization and low rank decomposition for the tensor coefficients. \citet{li2017parsimonious} assumed the Tucker factorization and imposed sparsity structures in both coefficients and responses.  \citet{lock2018tensor} also applied ridge penalty but they assumed low-rank CP decomposition on the tensor coefficients. \citet{raskutti2019convex} proposed a very general penalty term also related to lasso, group lasso, or other low rank regularizers that shrink many coefficients directly to zero. Moreover, these existing tensor regression models focused on either estimation or prediction, but this article focuses on variable selection.

We demonstrate through three simulation settings that the proposed TrimTenRidge approach achieves high success rates with well controlled false discoveries; it is robust for both sparse  (simulation setting 2) and non-sparse settings (simulations 1 and 3); and it has good extendability for more general settings. Specifically, simulation setting 1 is designed for tensor response-matrix predictor (imitating the motivating real data); Simulation setting 2 for scalar response-tensor predictor; and simulation setting 3 for tensor response-tensor predictor. We also applied the proposed TrimTenRidge approach to the human facial shape-GWAS data by modeling the entire facial shapes as a $2,342\times 7,160\times 3$ tensor.

The remainder of the paper is organized as follows: In Section 2 we elaborate on the details and theoretical properties. In Section 3 we assess the finite sample performance via numerical simulations. In Section 4 we implement real data analyses. Proofs are provided in the appendix.

\section{Methodology}
\subsection{Notation and preliminary}
We first introduce some notations following the format of \citet{kolda2009tensor}. The $order$ of a tensor is its number of dimensions. Let lowercase letters, e.g., $y$ to denote scalars; boldface lowercase letters, e.g., $\yv$ to denote vectors (tensors of order 1); boldface capital letters, e.g., $\Ybf$ to denote matrices (tensors of order 2), and Euler script letters, e.g., $\mathscr{Y}$ to denote higher-order tensors with order $\geq$ 3. Define $\Id_r$ as $r\times r$ identity matrix, $\0v_{m\times n}$ as $m\times n$ zero matrix, and $\|\cdot\|$ stands for the Frobenius norm. For two square matrices with same dimension $\Abf$ and $\Bbf$, $\Abf\leqslant \Bbf$ implies $\Bbf-\Abf$ is non-negative definite.

Given two tensors $\mathscr{A}\in \mathbb{R}^{I_1\times I_2\times\cdots\times I_M}$ and $\mathscr{B}\in \mathbb{R}^{J_1\times J_2\times \cdots\times J_N}$, the inner product over a common index $I_m=J_n=k$ is denoted as $\mathscr{C} =\langle \mathscr{A}, \mathscr{B}\rangle_{m,n}$, where each element of $\mathscr{C}$ is
\begin{equation*}
\mathscr{C}_{i_1\cdots i_{m-1}i_{m+1}\cdots i_Mj_1\cdots j_{n-1}j_{n+1}\cdots j_N} = \sum_{a=1}^k \mathscr{A}_{i_1\cdots i_{m-1} a i_{m+1}\cdots i_M}\mathscr{B}_{j_1\cdots j_{n-1} a j_{n+1}\cdots j_N}.
\end{equation*}
If there exist more than one common indices, one can calculate the inner product over several common indices. For example, for common indices $I_1 = J_3 = k_1$, $I_2 = J_4 = k_2$, denote $\mathscr{D} = \langle \mathscr{A}, \mathscr{B}\rangle_{(1,2),(3,4)}$, where each element of $\mathscr{D}$ is
\begin{equation*}
\mathscr{D}_{i_3 i_4 \cdots i_M j_1 j_2 j_5 \cdots j_N } = \sum_{a = 1}^{k_1} \sum_{b = 1}^{k_2} \mathscr{A}_{a b i_3 i_4\cdots i_M}\mathscr{B}_{j_1 j_2 a b j_5 \cdots j_N}.
\end{equation*}
The (full) inner product of two same-sized tensors $\mathscr{A},\mathscr{B}\in \mathbb{R}^{I_1\times I_2\times \cdots \times I_n}$ is defined as
\begin{equation*}
\langle \mathscr{A}, \mathscr{B}\rangle = \sum_{i_1=1}^{I_1}\sum_{i_2=1}^{I_2}\cdots\sum_{i_n=1}^{I_n}\mathscr{A}_{i_1 i_2 \dots i_n}\mathscr{B}_{i_1 i_2 \dots i_n},
\end{equation*}
which implies that $\langle \mathscr{A}, \mathscr{A}\rangle = \|\mathscr{A}\|^2$.

$Fibers$ are higher-order analogue of row or column vectors, which are defined as fixing all indexes of a tensor except for one dimension. For example, for a third-order tensor $\mathscr{Y}$, we use $\mathscr{Y}_{:jk}$, $\mathscr{Y}_{i:k}$ and $\mathscr{Y}_{ij:}$ to denote mode-1 (column) fibers, mode-2 (row) fibers and mode-3 (tube) fibers, respectively. $Slices$ are higher-order analogue of matrices, which are defined as fixing all indexes of a tensor except for two dimensions. For example, for a third-order tensor $\mathscr{Y}$, we use $\mathscr{Y}_{i::}$, $\mathscr{Y}_{:j:}$ and $\mathscr{Y}_{::k}$ to denote horizontal slices, lateral slices, and frontal slices, respectively.

\subsection{Ridge regression model for tensor response}
Consider the following linear tensor model
\begin{equation}\label{model1}
  \mathscr{Y} = \langle \Xbf, \mathscr{A}\rangle_{2,1} +\mathscr{E},
\end{equation}
where $\mathscr{Y}\in \mathbb{R}^{n\times d_1\times d_2}$, $\Xbf\in \mathbb{R}^{n\times p}$, $\mathscr{E}\in\mathbb{R}^{n\times d_1\times d_2}$, and $\mathscr{A}\in\mathbb{R}^{p\times d_1\times d_2}$. Here, $n$ is the number of observations, $p$ is the number of predictors, and $d_1, d_2$ are sizes of the tensor response. It is feasible to extend this model to a more challenging scenario when both response and predictors are higher order tensors ($>3$) utilizing similar ideas, however, in this article we only focus on this specific structure of model (\ref{model1}) because it is what the motivating data described in the real data analysis section requires.

Under ultra-high dimensional settings (i.e., $p = \exp[o(n^\xi)]$, $\xi>0$), $\mathscr{A}$ is generally not identifiable. Inspired by the idea of \citet{shao2012estimation} that was designed for a univariate response, we project $\mathscr{A}$ onto $\mathcal{R}(\Xbf)$, the linear space spanned by rows of $\Xbf$. Specifically, the SVD decomposition of $\Xbf$ yields $\Xbf = \Pbf\Dbf\Qbf^T$, where $\Pbf$ is an $n\times r$ matrix with $\Pbf^T\Pbf = \Id_r$, $\Qbf$ is a $p\times r$ matrix with $\Qbf^T\Qbf = \Id_r$, and $\Dbf$ is $r\times r$ diagonal matrix of full rank. Define $\Qbf_\perp$ as a $p\times (p-r)$ matrix such that $\Qbf^T\Qbf_\perp = \0v_{r\times (p-r)}$  and $\Qbf_\perp^T\Qbf_\perp = \Id_{p-r}$. Then it will be sufficient to transfer model (\ref{model1}) into the following model (2) by defining $\mathscr{B} = \langle\Qbf\Qbf^T, \mathscr{A}\rangle_{2,1}$ and $\mathscr{B}\in\mathbb{R}^{p\times d_1\times d_2}$,
\begin{equation}\label{model2}
  \mathscr{Y} = \langle \Xbf, \mathscr{B}\rangle_{2,1} +\mathscr{E}.
\end{equation}
Applying the ridge penalty to the model~(\ref{model2}), we estimate unknown parameters as
\begin{equation*}
  \hat{\mathscr{B}} = \langle(\Xbf^T \Xbf+h \Id_p)^{-1}\Xbf^T, \mathscr{Y}\rangle_{2,1},
\end{equation*}
where $h$ is the tuning parameter for ridge penalty.

Suppose $\mathscr{E}_{: ij}|\Xbf$ follows a sub-Gaussian distribution with variance proxy $\sigma^2$, where $i=1,2,\dots, d_1$, $j=1,2,\dots,d_2$. It follows that
\begin{equation*}
  \begin{split}
     \mathbb{E}[bias(\hat{\mathscr{B}})|\Xbf] & = \mathbb{E} [\hat{\mathscr{B}}|\Xbf]-\mathscr{B} \\
       & =\langle (\Xbf^T\Xbf+h\Id_{p})^{-1} \Xbf^T, \langle \Xbf, \mathscr{B}\rangle \rangle_{2,1} -\mathscr{B} \\
       & =\langle (\Xbf^T\Xbf+h\Id_{p})^{-1}(-h), \mathscr{B}\rangle_{2,1} \\
       & =-\langle \frac{1}{h} \Xbf^T\Xbf+\Id_{p})^{-1}, \mathscr{B}\rangle_{2,1} \\
       & =\langle -\Qbf(\frac{1}{h}\Dbf^2+\Id_r)^{-1}\Qbf^T, \mathscr{B}\rangle_{2,1}.
  \end{split}
\end{equation*}
It also yields that
\begin{equation*}
\begin{split}
\Var[\hat{\mathscr{B}}_{: ij}|\Xbf] & \leqslant \sigma^2 (\Xbf^T\Xbf+h\Id_p)^{-1}\Xbf^T\Xbf(\Xbf^T\Xbf+h\Id_p)^{-1}\\
& \leqslant \sigma^2(\Xbf^T\Xbf+h\Id_p)^{-1} \\
& \leqslant \sigma^2h^{-1}\Id_p,
\end{split}
\end{equation*}
for $i=1,2,\dots, d_1$, $j=1,2,\dots,d_2$.

Before claiming the theoretical statements, we first specify the two conditions that we need to assume:

  \begin{itemize}
  \item Condition (C1): Let  $\lambda_1$ be the smallest positive eigenvalue of $\Xbf^T\Xbf$. We assume that
  \begin{equation}
  \lambda_1^{-1} = O_P(n^{-\eta}), \qquad\eta\leqslant 1.
\end{equation}
  \item Condition (C2): We also assume that $\mathscr{B}$ is upper bounded, that is, for each individual response $\mathscr{Y}_{ij}$, $i=1,2,\dots,d_1$, $j=1,2,\dots,d_2$, the $p$-dimensional coefficient fiber, $\mathscr{B}_{: ij} = (\mathscr{B}_{1ij},\mathscr{B}_{2ij},\dots, \mathscr{B}_{pij})$, satisfies $\|\mathscr{B}_{: ij}\| = O(n^\tau)$, and hence
\begin{equation}
  \|\mathscr{B}\| = \sqrt{d_1d_2}O(n^\tau),\qquad \textrm{for some $0<$}\tau<\eta.
\end{equation}
\end{itemize}
Note that condition (C2) naturally holds if the number of nonzero components of $\mathscr{A}$ is $O(n^{2\tau})$ and all components of it are bounded by a constant $M$ since $\|\mathscr{B}\|\leqslant \|\mathscr{A}\| \leqslant M n^\tau$. We want to emphasize that we do not require any sparsity assumption, e.g., the condition (C2) still holds without any slice-wise or fiber-wise sparse assumptions on $\mathscr{B}$.

\begin{lemma}\label{lemma1}
 Consider the model (2) and assume that conditions (C1) and (C2) hold. We have the following two conclusions: As $n\to\infty$,
  \begin{enumerate}
  \item $\mathbb{E}[(\langle \mathscr{G}, \hat{\mathscr{B}}\rangle - \langle \mathscr{G}, \mathscr{B}\rangle)^2 | \Xbf] = d_1d_2O(h^{-1})+d_1d_2O(h^2n^{-2(\eta-\tau)})$ for $\mathscr{G}\in \mathbb{R}^{p\times d_1\times d_2}$ and $\|\mathscr{G}\| = 1$;
  \item $n^{-1}\mathbb{E}[\|\langle \Xbf, \hat{\mathscr{B}}\rangle_{2,1} -\langle \Xbf, \mathscr{B}\rangle_{2,1} \|^2 |\Xbf] = d_1d_2\sigma^2O(rn^{-1})+d_1d_2O(h^2n^{-(1+\eta-2\tau)})$.
\end{enumerate}
\end{lemma}
\subsection{Variable selection by trimming the ridge regression estimator}
Unlike the Lasso penalty, the ridge estimator can not shrink unimportant predictors directly to zero, which may not accommodate the needs of real data analysis if it has some sparse structure on $\mathscr{B}$. We introduce a trimming approach to facilitate the variable selection process for the model (\ref{model2}) \citep{shao2012estimation}. Specifically, we define a threshold value $a_n = Cn^{-\alpha}$ where $C>0$, $\alpha>0$, and define the trimmed ridge estimator $\tilde{\mathscr{B}}$ as
\begin{equation}\label{truncate}
\widetilde{\mathscr{B}}_{kij} = \begin{cases}
\hat{\mathscr{B}}_{kij}, & \quad \text{if} ~ |\hat{\mathscr{B}}_{kij}| > a_n, \\
0, & \quad \text{otherwise},
\end{cases}
\end{equation}
for $k=1,2,\dots,p$, $i=1,2,\dots,d_1$, and $j=1,2,\dots,d_2$.

\begin{theorem}\label{theorem1}
 Consider the model (2) and assume that the error term $\mathscr{E}_{: ij}|\Xbf; i=1,2,\dots, d_1, j=1,2,\dots,d_2$ follows an i.i.d. sub-Gaussian distribution with variance proxy $\sigma^2$, meanwhile conditions (C1) and (C2) hold. Let $u_n = 1+(\log\log n)^{-1}$ and $h = M_2a_n^{-2}(\log\log n) n^\theta$ where $M_2>0$ is a constant. Let $a_n$ be the threshold value and $0<\alpha < (\eta-\tau-\theta)/3$. Then we have
  \begin{equation*}
    P(\mathcal{M}_{\mathscr{B},a_nu_n}\subset \mathcal{M}_{\hat{\mathscr{B}},a_n} \subset \mathcal{M}_{\mathscr{B},a_n/u_n}) \geqslant 1 - 4p\exp \{-\frac{c_1n^{2\theta+2\alpha}}{2d_1^2d_2^2\sigma^2}\},
  \end{equation*}
  where $\mathcal{M}_{\mathscr{B}, a_nu_n}$ denotes the set of indices of components of $\mathscr{B}$ whose absolute values are greater than $a_nu_n$.
\end{theorem}
Theorem~\ref{theorem1} guarantees that this trimmed ridge estimator preserves selection consistency, and it can handle ultrahigh dimensionality of order $\log(p) = o(n^{2\theta+2\alpha})$.

\zhao{Now if one does want to have a sparse situation, the following corollary, as a natural extension of Theorem 1, still works to prove the selection consistency}. Define $\mathcal{I}_*=\{(k,i,j): \mathscr{B}_{kij}\neq0\}$ and $\widehat{\mathcal{I}}=\{(k,i,j): \widetilde{\mathcal{B}}_{kij}\neq0\}$. Let $m_n=\sum_{(i,j,k)\in\mathcal{I}_*}|\mathscr{B}_{kij}|$.

\begin{corollary}
If condition (C1) and $m_n=\sqrt{d_1d_2}O(n^\tau)$ hold, we have $P(\widehat{\mathcal{I}}=\mathcal{I}_*)\to1$.
\end{corollary}
\subsection{Tuning parameter setup}
There are two tuning parameters involved in the Equation (5), the ridge penalty parameter $h$ and the threshold parameter $a_n$. We will apply the generalized cross-validation (GCV) to choose $h$ and use cross-validation to choose $a_n$ by minimizing the prediction mean squared error.

For ultrahigh dimensional settings, the number of predictors are much larger than the sample size, i.e., $p>>n$. Since the ridge estimator $\hat{\mathscr{B}} = \langle(\Xbf^T\Xbf+h\Id_p)^{-1}\Xbf^T, \mathscr{Y}\rangle_{2,1}$ requires the inverse of a $p\times p$ matrix, it imposes significant computational and memory challenges for a large $p$, which is the case of our motivating example. Based on the fact that \citep{shao2012estimation}
\begin{equation*}
(\Xbf^T\Xbf+h\Id_p)^{-1}\Xbf^T = \Xbf^T(\Xbf\Xbf^T+h\Id_n)^{-1},
\end{equation*}
 we instead compute the inverse of an $n\times n$ matrix, which greatly reduce the computational cost compared to the inverse of the original $p\times p$ matrix.

Define the hat matrix $\Abf(h)$ as
\begin{equation*}
\Abf(h) = \Xbf\Xbf^T(\Xbf\Xbf^T+h\Id_n)^{-1}.
\end{equation*}
Then the prediction error of the generalized cross-validation can be computed as
\begin{equation*}
V(h) = \frac{\frac{1}{n}\|\langle\Id_n-\Abf(h),\mathscr{Y}\rangle_{2,1}\|^2}{[\frac{1}{n}trace (\Id_n-\Abf(h))]^2}.
\end{equation*}
Finally, we choose $h$ from
\begin{equation*}
\hat{h} = \argmin_{h\in\mathbb{R}^+} V(h).
\end{equation*}
In the following we summarize the detailed scheme for the two tuning parameters setup procedure:
\begin{itemize}
\item Step 1: Split the data into training set and validation set, denoted as $\Xbf_{tr}$, $\mathscr{Y}_{tr}$, $\Xbf_{va}$ and $\mathscr{Y}_{va}$.
\item Step 2: Calculate the generalized cross-validation estimate of $h$ using the entire training data, denoted as $h_{GCV}$.
\item Step 3: Calculate $\hat{\mathscr{B}}=\langle \Xbf_{tr}^T(\Xbf_{tr}\Xbf_{tr}^T+h_{GCV}\Id_n)^{-1}, \mathscr{Y}_{tr}\rangle_{2,1}$.
\item Step 4: Give a set of candidate $a_n$'s with a fine scale (denoted as $S_a$), and then apply equation (\ref{truncate}) to calculate a sequence of $\widetilde{\mathscr{B}}$'s for each of the $a_n$.
\item Step 5: Calculate prediction mean squared errors $MSE = \|\mathscr{Y}_{va} - \langle\Xbf_{va},\widetilde{\mathscr{B}}\rangle_{2,1}\|^2$ for each of the $a_n$'s, and locate the optimal choice of $\hat{a}_n$ by minimizing the MSE as follows,
\begin{equation*}
\hat{a}_n = \argmin_{a_n\in S_a}MSE.
\end{equation*}
\item Step 6: Repeat the steps 1-5 for 100 times, finally locate the mean value of $h_{GCV}$ and $\hat{a}_n$ across 100 replications.
\end{itemize}

\section{Numerical Studies}
In this section we assess the performance of the proposed TrimTenRidge approach through three simulation settings under both sparse and non-sparse scenarios. Specifically,  Simulation 2 represents a sparse setting because its noise entries are all zero, and Simulations 1 and 3 represent non-sparse settings because their noise entries are all generated from Uniform(0,0.001), which represent a scenario that a large amount of noise still have nonzero but weak effects to mimic real data complexity.

Each simulation setting is replicated 100 times and two boxplots are made to demonstrate the sensitivity and specificity of the method. For sensitivity, we report the true positive rate, where 1 implies that we correctly select all true variables and 0 means that no true variable is selected. For 1-specificity, we report the false negative rate, where 1 implies that we select all noise variables and 0 means that no noise variable is selected. In summary, we want a high sensitivity (closer to 1) and a high specificity (i.e., 1-specificity closer to 0).

\subsection{Simulation setting 1: tensor response-matrix predictor scenario}
 To mimic the motivating shape-GWAS data that has all predictors forming a matrix and responses in the format of tensor,  we generate data exactly from the model (\ref{model2}). We fix $(d_1,d_2) = (4,3)$ and investigate nine combinations by varying $n=200,500,1000$ and $p=2000,5000,10000$. We generate $\Xbf$ from a multivariate normal distribution with mean $\0v$ and covariance matrix $\Sigmav$, where $\Sigmav_{i,j} = 0.8^{|i-j|}$ for $1\leqslant i,j \leqslant p$. We generate $\mathscr{B}_{1,1,1}, \mathscr{B}_{2,1,1}, \mathscr{B}_{1,3,4}, \mathscr{B}_{2,2,2}, \mathscr{B}_{12,1,2}, \mathscr{B}_{12,2,3}$, and $\mathscr{B}_{22,3,3}$ from $\mathrm{Uniform}(1,2)$ as the seven true coefficients. All remaining $p\times d_1\times d_2-7$ entries of $\mathscr{B}$ are generated from $\mathrm{Uniform}(0,0.001)$ as noise coefficients. Note that those noise coefficients are not exactly 0, hence, $\mathscr{B}$ in the Simulation setting 1 is not sparse. Additionally, the error term $\mathscr{E}$ are generate from i.i.d. standard normal.

\begin{figure}[H]
    \centering
    \begin{minipage}{0.49\textwidth}
    
        \includegraphics[width = 0.9\textwidth]{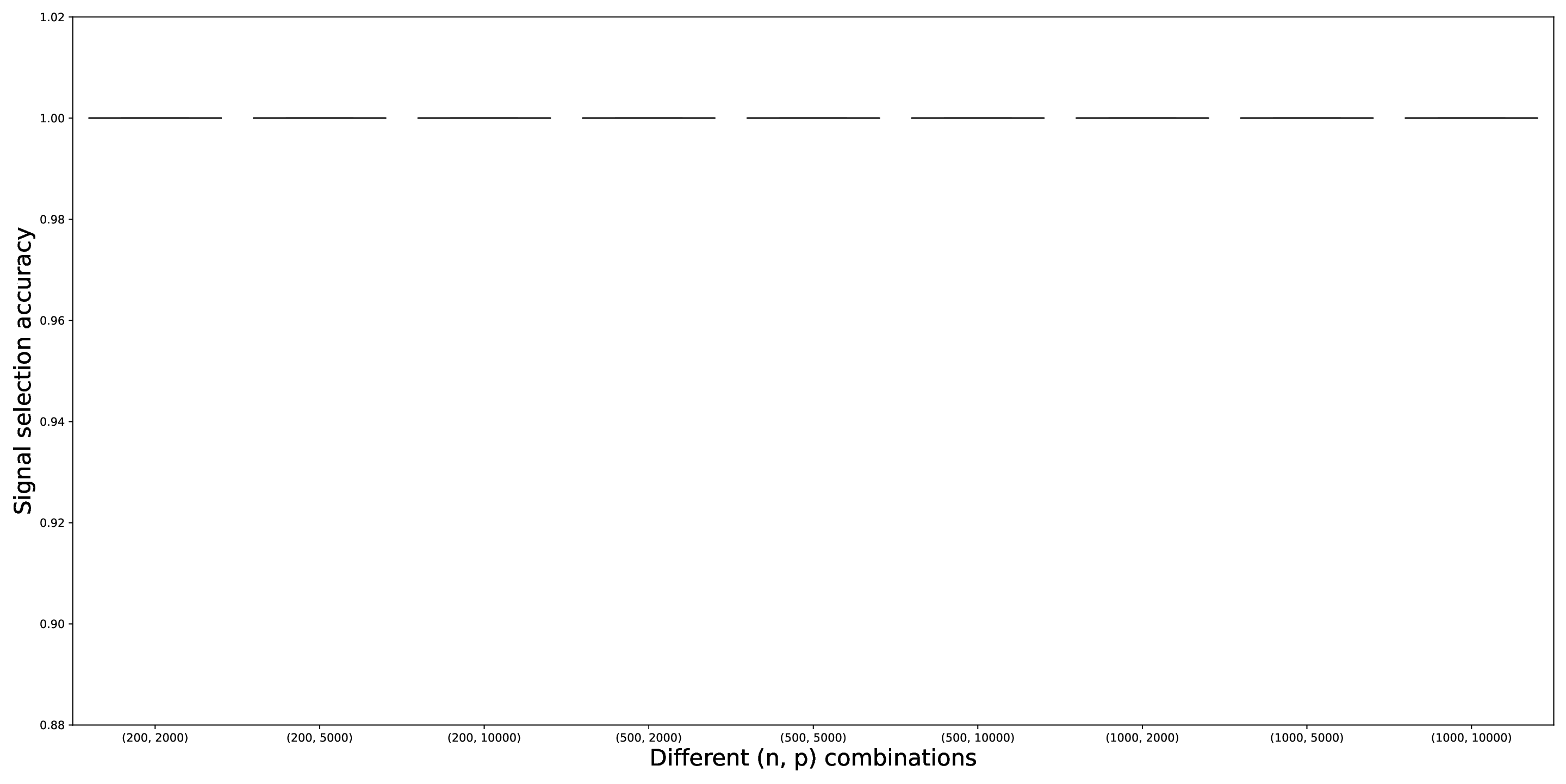}
        \captionof{figure}{The true positive rates of TrimTenRidge approach for Simulation Setting 1 when assessing if it successfully selects all the seven true coefficients.}
        \label{true1}
    \end{minipage}
    \hfill
    \begin{minipage}{0.49\textwidth}
        
        \includegraphics[width = 0.9\textwidth]{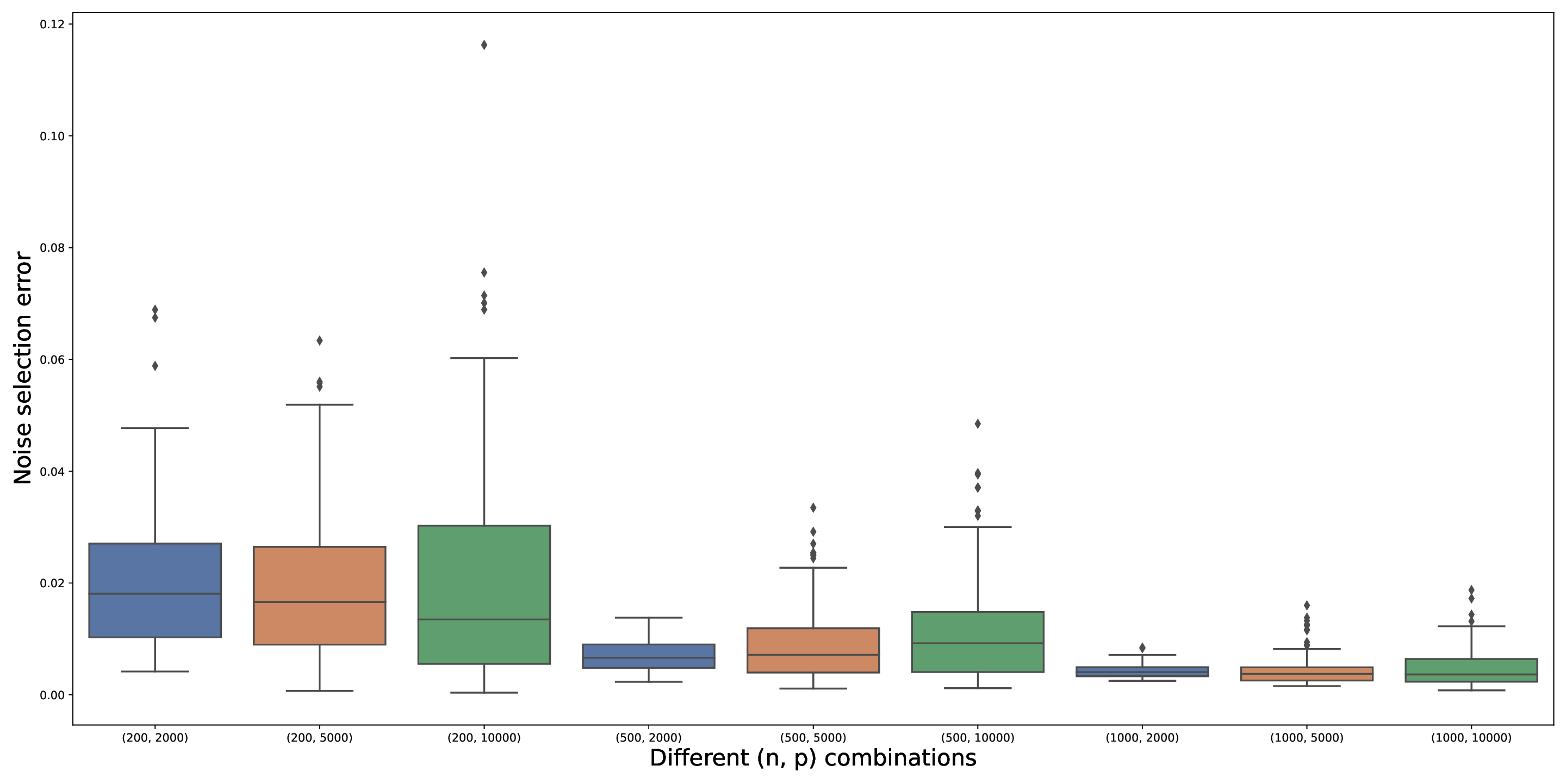}
        \captionof{figure}{The false negative rates of TrimTenRidge approach for Simulation Setting 1 when assessing if it wrongly selects any of the $p\times d_1\times d_2 -7$ noise coefficients.}
        \label{error1}
    \end{minipage}
\end{figure}
As demonstrated in Fig. \ref{true1}, the TrimTenRidge approach achieves 100\% accuracy and successfully identifies all of the 7 true coefficients in all of the nine ($n$, $p$) combinations without missing any true signal. We notice from Fig. \ref{error1} that the false negative rates consistently decrease as $n$ increases but increase as $p$ increase for each fixed $n$, which meet the theoretical expectations in general. We conclude that the false negative rates are below 0.05 for the six combinations when $n >500$. The worst case comes from $n=200$, where the false negative rates are still below 0.08, which is acceptable given that it has $120,000 (=10,000\times4\times3)$ coefficients to estimate and select from.

\subsection{Simulation setting 2: univariate response-tensor predictor scenario}
In this simulation, we generate data from
\begin{equation*}
    \yv = \langle \mathscr{X}, \mathscr{B}\rangle_{(2,3,4),(1,2,3)} + \ev,
\end{equation*}
where $\yv,\ev\in\mathbb{R}^{n}$, $\mathscr{X}\in\mathbb{R}^{n\times p\times p\times p}$ and $\mathscr{B}\in\mathbb{R}^{p\times p\times p}$. Here, $p\times p\times p$ is designed for dimensions of the tensor predictor. This simulation is very useful in some practices. For example, multi-omics data is usually collected from multiple platforms and they can have the same dimension with matched genes.  We also consider nine combinations by varying $n=200,500,1000$ and $p = 50,80,100$. Note that it is a very challenging scenario. For example, even for $p=50$ the total number of coefficients is already $50^3=125,000$. We generate each entry of $\mathscr{X}$ from i.i.d. standard normal distribution. For coefficient $\mathscr{B}$, we generate $\mathscr{B}_{1,11,1}, \mathscr{B}_{21,3,14}, \mathscr{B}_{11,11,6},$ and $\mathscr{B}_{16,31,21}$ from uniform (2,4) as four true signals and all remaining $p\times p\times p-4$ entries of $\mathscr{B}$ are directly set to be 0 as noise. Note that those noise coefficients are exactly 0, hence, $\mathscr{B}$ in the Simulation setting 2 is sparse. The error term $\ev$ is also generated from i.i.d. standard normal distribution.

\begin{figure}[H]
    \centering
    \begin{minipage}{0.49\textwidth}
        \includegraphics[width = 0.9\textwidth]{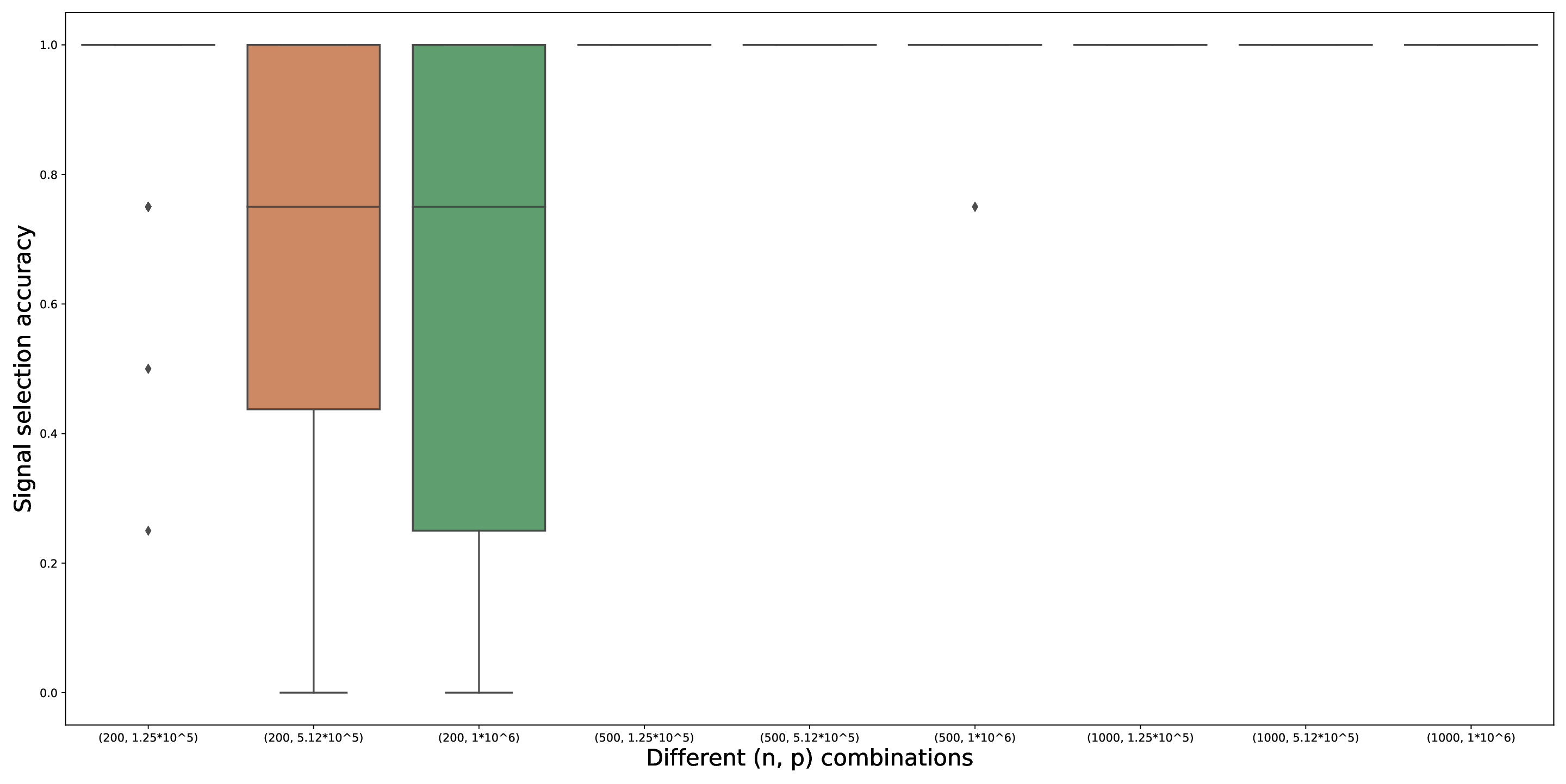}
        \captionof{figure}{The true positive rates of TrimTenRidge approach for Simulation Setting 2 when assessing if it successfully selects all the $4$ true coefficients.}
        \label{true2}
    \end{minipage}
    \hfill
    \begin{minipage}{0.49\textwidth}
        \includegraphics[width = 0.9\textwidth]{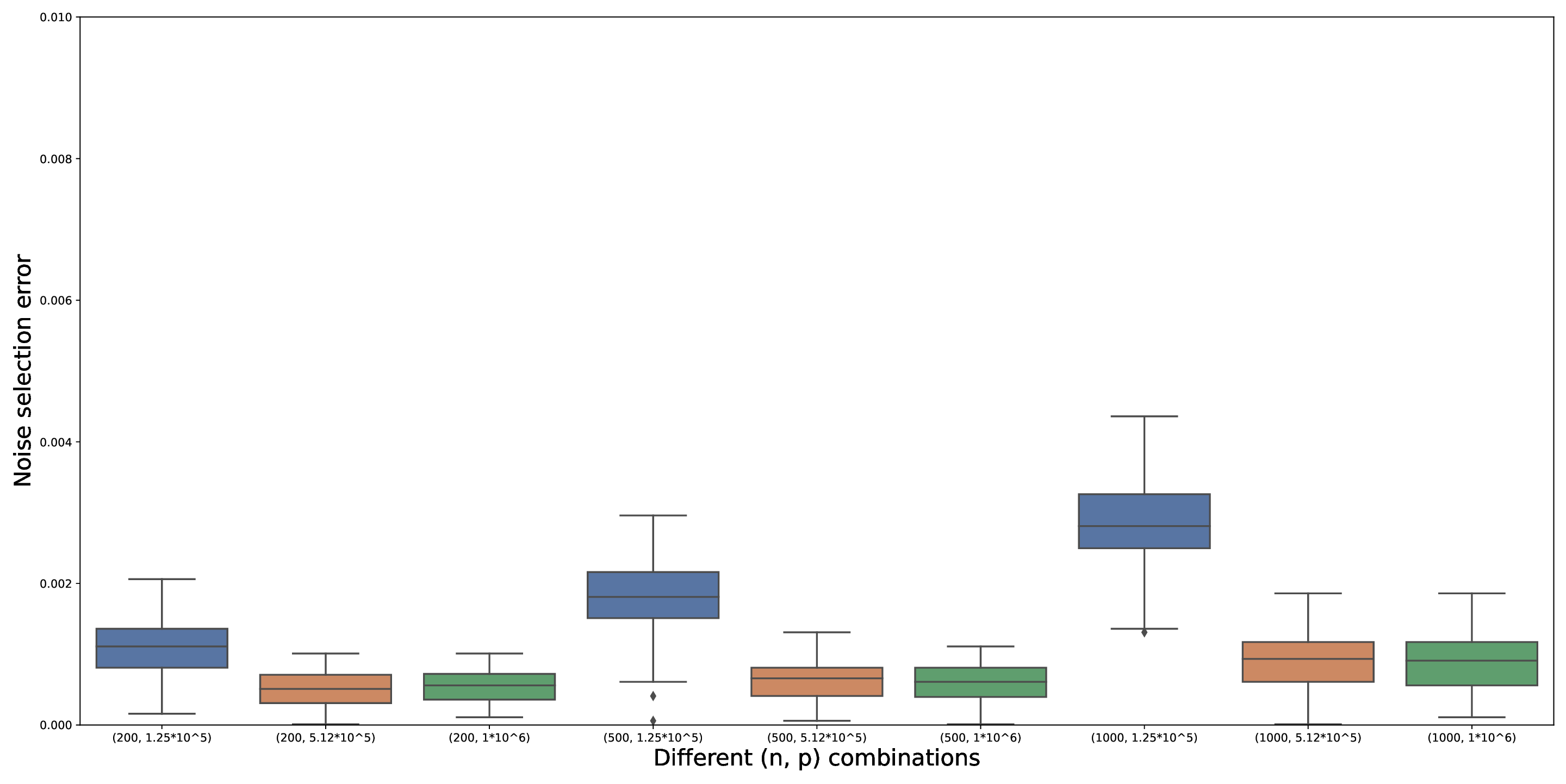}
        \captionof{figure}{The false negative rates of TrimTenRidge approach for Simulation Setting 2 when assessing if it wrongly selects any of the $p^3 -4$ noise coefficients.}
        \label{error2}
    \end{minipage}
\end{figure}
From Fig. \ref{true2}, we can see that the true positive rates are only 75\% or 50\% when $n=200$. It is not surprising given a number of $100^3$ coefficients with only sample size of 200. When we increase $n$ to be 500, the true positive rates consistently increase back to 100\%. Fig. \ref{error2} demonstrates that the false negative rates of noise coefficients are consistently below 0.005, which is quite impressive.

\subsection{Simulation setting 3: tensor response-tensor predictor scenario}
In this simulation, we consider the tensor response and tensor predictor to further extend the scope of the approach to a more general setting. We generate data from the following model
\begin{equation*}
    \mathscr{Y} = \langle \mathscr{X},\mathscr{B}\rangle_{(2,3), (1,2)} + \mathscr{E},
\end{equation*}
where $\mathscr{Y},\mathscr{E}\in\mathbb{R}^{n\times d_1\times d_2}$, $\mathscr{X}\in\mathbb{R}^{n\times p_1\times p_2}$ and $\mathscr{B}\in\mathbb{R}^{p_1\times p_2\times d_1\times d_2}$, $p_1, p_2$ are the sizes of the predictors, and $d_1,d_2$ are the sizes of the responses. We fix $(d_1,d_2,p_2) = (100,10,4)$ and investigate nine combinations by varying $n=1000,2000,3000$ and $p_1 = 2000,5000,10000$. We generate each entry of $\mathscr{X}$ from i.i.d. $\mathrm{Uniform}(-1,1)$. For the coefficient tensor, we generate $\mathscr{B}_{1,1-4,1,1}, \mathscr{B}_{101,3,51,1-10}, \mathscr{B}_{1001, 4,21,6}$ from $\mathrm{Uniform}(1,2)$ as fifteen true signals and all remaining $p_1\times p_2\times d_1\times d_2-15$ entries of $\mathscr{B}$ from $\mathrm{Uniform}(0,0.001)$ as noise. Again this simulation represents a non-sparse setting. The error term $\mathscr{E}_{:ij}$ are generated from i.i.d. standard normal distribution.

\begin{figure}[H]
    \centering
    \begin{minipage}[t][][b]{0.49\textwidth}
        \includegraphics[width = 0.9\textwidth]{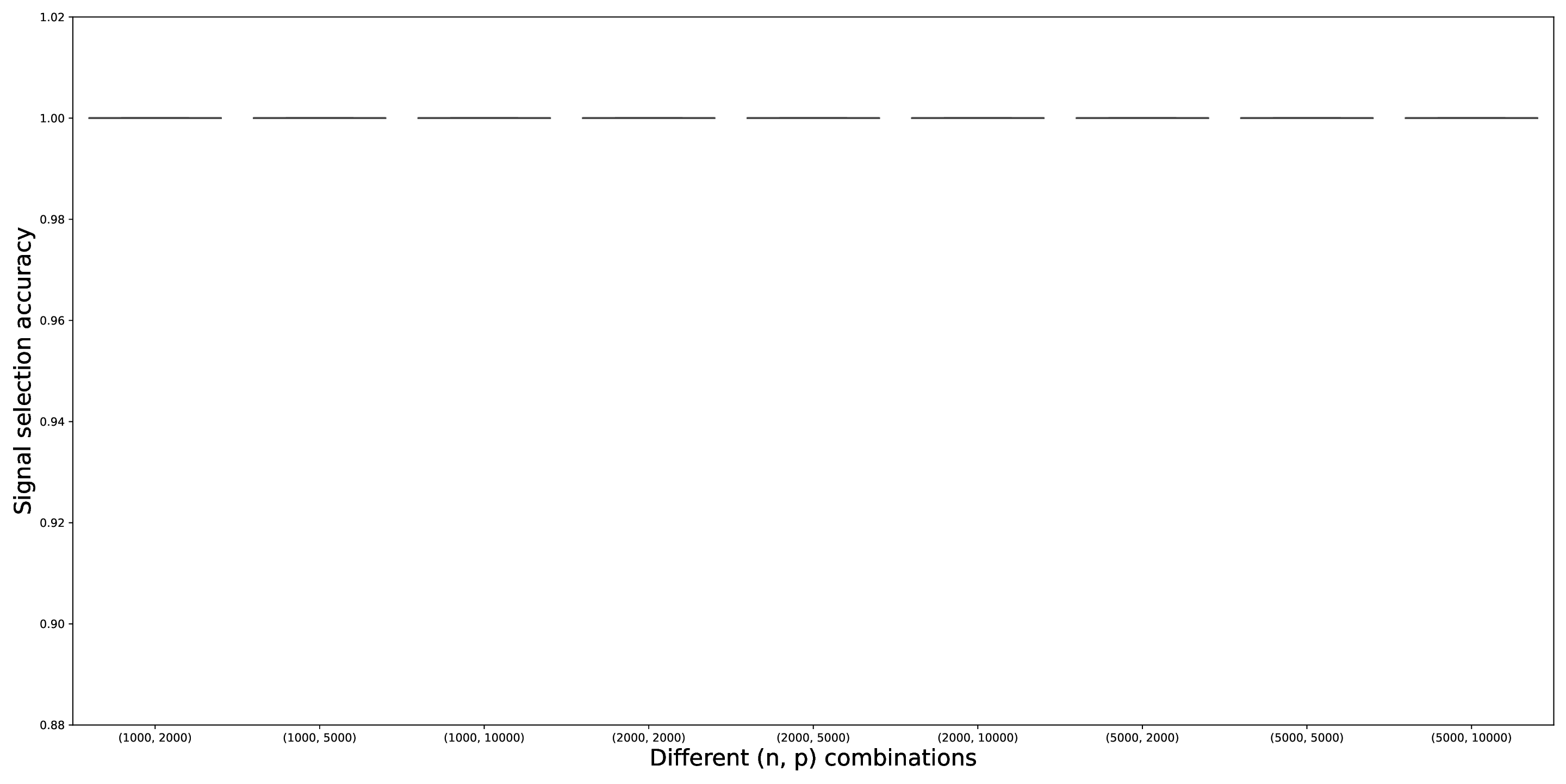}
        \captionof{figure}{The true positive rates of TrimTenRidge approach for Simulation Setting 3 when assessing if it successfully selects all the $15$ true coefficients..}
        \label{true3}
    \end{minipage}
    \hfill
    \begin{minipage}[t][][b]{0.49\textwidth}
        \includegraphics[width = 0.9\textwidth]{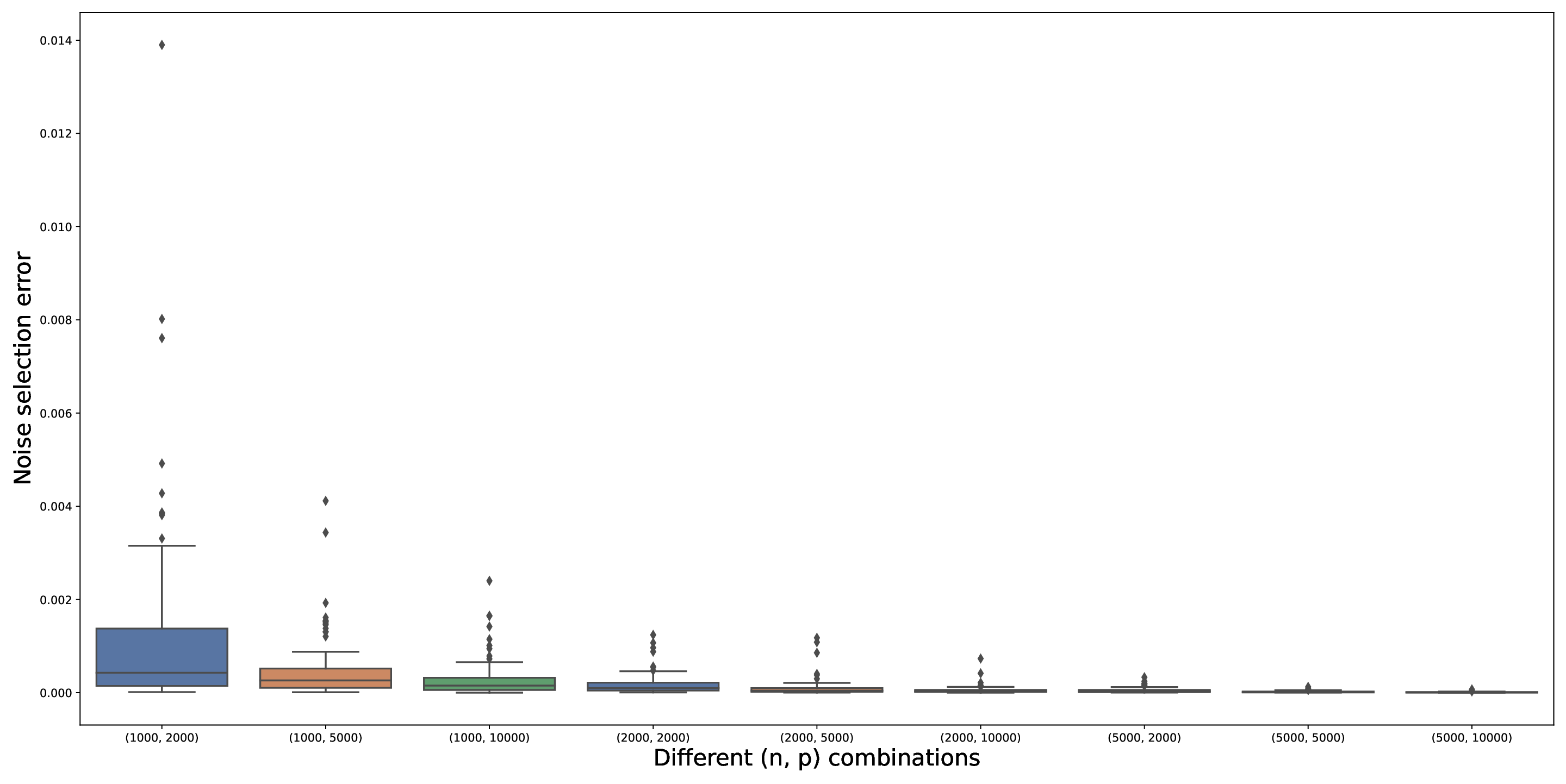}
        \captionof{figure}{The false negative rates of TrimTenRidge approach for Simulation Setting 3 when assessing if it wrongly selects any of the $p_1\times p_2\times d_1\times d_2-15$ noise coefficients.}
        \label{error3}
    \end{minipage}
\end{figure}
As demonstrated in Fig. \ref{true3}, the 15 true coefficients are 100\% selected by the TrimTenRidge approach for all the nine combinations.  Fig. \ref{error3} shows that the false negative rates are well controlled below 0.01 for all the nine combinations with a significantly decreasing trend as $n$ increases.

\section{Real Data Analysis}
The human facial shape-GWAS cohort, representing an example of valuable ultrahigh dimensional big data, has received a lot of attention \citep{claes2018genome, liu2021genome, kang2017manifold,hoskens20213d, bonfante2021gwas, xiong2019novel, white2021insights}. Pioneering research published recently has brought significant breakthroughs in phenotyping, using data driven approaches to extract shape information from 3D facial surface scans \citep{claes2018genome}, assembling multiple datasets to achieve large sample sizes \citep{white2021insights}, exploring different ancestries \citep{liu2021genome}, and investigating biological interpretations and functional annotations for their findings \citep{naqvi2021shared}. However, the statistical methods used to uncover SNP­-shape association still have room for improvement.

For example, the aforementioned studies first decreased the dimensionality of the shape vector by principal component analysis (PCA) and extracted relevant PCs, and then applied multivariate canonical correlation analysis (CCA) to detect significant SNPs associated with the extracted PCs (the PCs are modeled as the response) \citep{claes2018genome, liu2021genome, white2021insights}. Other works have applied multivariate linear mixed models to compute p-values for each SNP \citep{bonfante2021gwas, xiong2019novel}.  The Multivariate linear mixed models or the multivariate CCA work well for modeling shape as multidimensional traits. However, they are still restricted to testing each SNP in isolation and do not consider the joint effects of other SNPs; additionally, modeling the PCs as the response variables may not be as easy to interpret as using the original shape data.
 As noted by Atwell, “At least for complex traits, the problem is better thought of as model misspec­ification: when we carry out GWAS using a single SNP at a time (as is done in \citet{atwell2010genome} and in most other previous GWAS), we are in effect modeling a multifactorial trait as if it were due to a single locus. The polygenic background of the trait is ignored.” Carlsen et al. demonstrated through 48 simulation settings that a single­-SNP model, like the Cochran-Armitage (CA) trend test, yielded both high false­ positives and high false­ negatives \citep{carlsen2016exploiting}.

In this section, we explore human facial shape-GWAS data by leveraging the proposed TrimTenRidge approach to model shape as a tensor. This approach considers the joint effects of all SNPs in one model simultaneously instead of screening each SNP one by one. However, since the total number of SNPs (9, 478, 608) is too large, we performed the selection process for each of the 23 chromosomes separately.

\subsection{Data information and pre-processing}

\begin{itemize}

    \item \textbf{The response}: Digital stereophotogrammetry was utilized to obtain a 3D facial image for each of the 2,342 participants\citep{snyders2014development, weinberg20163d}. Then a dense correspondence alignment was performed for all the 3D facial images to establish homology of 7,160 quasi-landmark points \citep{claes2012improved, claes2014modeling}. Since each point consists of the XYZ coordinates, we model the facial shape response as a $2,342\times 7,160\times 3$ tensor structure.

    \item \textbf{The predictors}: The 2,342 participants were genotyped using the Illumina OminExpress + Exome v1.2 array. Then the SHAPEIT2 was utilized to obtain pre-phasing haplotypes \citep{delaneau2013improved}, and imputation was performed using IMPUTE2 \citep{howie2009flexible}, with the 1000 Genomes Project Phase 3 as the reference panel \citep{10002015global}. We also applied standard quality control filters to pre-process the genomic data and finally retain 6,322,724 SNPs. Specifically, SNP-level (INFO score $>0.5$) and genotype-per-participant-level (genotype probability $>0.9$) filters were used to omit poorly-imputed variants; After imputation, we further remove SNPs with either missing values or having the same values across all subjects. Following the steps of \citet{claes2018genome}, we minimized confounding factors caused by population structure by extracting four PCs using PCA of approximately 97,000 autosomal genotyped SNPs chosen for call rate ($>$95\%), MAF ($>$0.05) and pairwise $r^2$ ($<0.1$ across variants in a sliding window of 10Mb). In addition, we also added the sex variable into the model. Altogether the predictor data forms a $2,342\times 6,322,729$ matrix.

\end{itemize}

\begin{figure}[H]
    \centering
    \begin{minipage}{0.45\textwidth}
        \includegraphics[width = 0.99\textwidth]{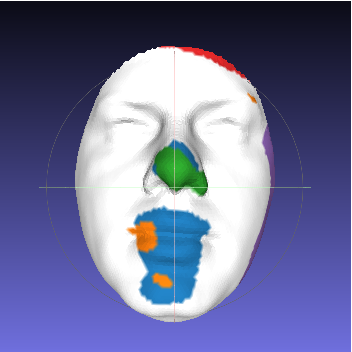}
        \captionof{figure}{The affected regions of five significant SNPs on chromosome 3 (\emph{rs6779419, rs7643249}, etc.)}
        \label{ch3}
    \end{minipage}
	\hfill
    \begin{minipage}{0.45\textwidth}
        \includegraphics[width = 0.99\textwidth]{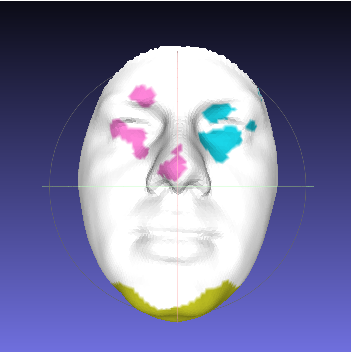}
        \captionof{figure}{The affected regions of three significant SNPs on chromosome 7 (\emph{rs587741, rs17657924}, etc.)}
        \label{ch7}
    \end{minipage}
\end{figure}

\begin{figure}[H]
    \centering
    \begin{minipage}{0.45\textwidth}
        \includegraphics[width = 0.99\textwidth]{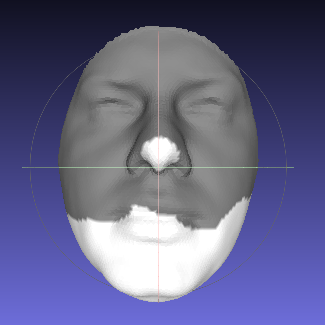}
        \captionof{figure}{The affected region of SNP \emph{rs4980297} on chromosome 10}
        \label{ch10}
    \end{minipage}
    \hfill
    \begin{minipage}{0.45\textwidth}
        \includegraphics[width = 0.99\textwidth]{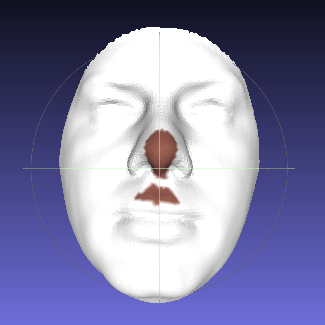}
        \captionof{figure}{The affected region of SNP \emph{rs4675833} on chromosome 2}
        \label{ch2}
    \end{minipage}
\end{figure}

After applying the proposed TrimTenRidge approach to this facial shape-GWAS dataset, we were able to detect 2,391 SNPs with nonzero coefficients. We annotated genes near these 2,391 SNPs utilizing PLINK 1.9 \citep{chang2015second} with SNP attribute file \emph{snp129.attrib.gz} and gene list file \emph{glist-hg38}. In Table \ref{gene_annotation}, we list twelve representative findings and report the detected SNPs, their corresponding gene symbols, and the corresponding facial regions that are associated with the selected genes (differentiated by colors). The detailed information for all the 2,391 detected SNPs can be found in the supplementary file. Note, this list includes several well-established craniofacial genes, for example, \emph{ALX4}, \emph{BMP2}, and \emph{BMP7}. From the last two columns of Table \ref{gene_annotation}, we want to emphasize again that the TrimTenRidge approach not only detects genes associated with local facial regions, such as eyes, lips, noses, etc; but also detects genes that are associated with the entire face. In particular, the two most important SNPs that are ranked the highest by the TrimTenRidge approach, \emph{rs6109993} (gene \emph{TASP1}) and \emph{rs1479927}(gene \emph{ASB11}), are SNPs that impact the entire face.

Among the twelve results summarized in Table \ref{gene_annotation}, three genes are confirmed by other works in the literature using different datasets, foci, and approaches. Specifically, SNP \emph{rs6109993} (gene \emph{TASP1}) was found to be related to chin dimples \citep{pickrell2016detection,white2021insights};  \citet{boonsawat2019elucidation} found that the variants of gene \emph{ASB11} may contribute to risk in microcephaly. We found that \emph{rs587741} (gene \emph{SRPK2}) is associated with the eye and nose tip region. \citet{nevado2014new} showed that the deletion of 7q22.2-q22.3 (including gene \emph{SRPK2, MLL5, RINT1} and \emph{LHFPL3}) may results in facial dysmorphology.

In addition to confirming three aforementioned genes that have already been reported by the literature, we also detect some novel findings. We visually demonstrate our new findings in Figures 7-10. After plotting all non-zero coefficients of the SNP \emph{rs7617493} (\emph{IL5RA}) in \textcolor{myblue}{blue} color, we notice that the associated facial region for this gene is concentrated in the lip and chin area, as well as the nasal tip and nostrils. By making similar plots for all other important SNPs on chromosome 3 from Table \ref{gene_annotation}, we find that the effects of these SNPs seem mainly concentrated on the nose, mouth, forehead, and lateral edge of the face (see Fig. \ref{ch3}). Specifically, all non-zero coefficients of the SNP \emph{rs7643249}  (near to \emph{LRRC34}) are located around the alae and the tip of the nose (\textcolor{mygreen}{green}); the SNP \emph{rs2080794} (near \emph{CLDN1}) is associated with forehead shape (\textcolor{myred}{red}); the SNP \emph{rs6771833} (near \emph{CNTN6}) is related with the left ear and lateral facial regions (\textcolor{mypurple}{purple}); and the SNP \emph{rs6779419} (\emph{SUCLG2-AS1}) impacts the lip and chin regions (\textcolor{myorange}{orange}). Moreover, Fig. \ref{ch7} demonstrates the effects of all important SNPs on chromosome 7 from Table \ref{gene_annotation}. The effects of these SNPs seem to mainly impact the eyes and lower mandibular regions. Specifically, \emph{rs587741} (\emph{SRPK2}) is associated with the right eye, eye brow, and nose regions (\textcolor{mypink}{pink}) ; \emph{rs17657924} (\emph{LOC100506136}) is related with chin and mandibular regions (\textcolor{myyellow}{yellow}); and the effects of \emph{rs847375} (near \emph{AGR3}) are focused on the left eye, eye brow, and nose regions (\textcolor{mycyan}{cyan}).

Furthermore, SNP \emph{rs4980297} (near \emph{CTBP2}) from chromosome 10 (Fig. \ref{ch10}) has a strong association with the entire upper half of the face, except the tip of the nose (\textcolor{mygray}{gray}). On the contrary, the SNP \emph{rs4675833} (\emph{LINC01237}) from chromosome 2 (Fig. \ref{ch2}) is related to only a small region involving the tip of the nose and upper lip (\textcolor{mylotus}{lotus}).

\begin{table}
\centering
\caption{Selected SNPs and their affecting regions}
    \label{gene_annotation}
\resizebox{\textwidth}{!}{%
\begin{tabular}{cccccccc}
\toprule
Chromosome & SNP & Position & Locus & MAF & Candidate gene (distance) & Effect & Color \\
\midrule
20 & rs6109993 & 13635155 & 20p12.1 & 0.3940 & TASP1 (0) & whole face & N.A. \\
X & rs1479927 & 15282822 & Xp22.2 & 0.4023 & ASB11 (0) & whole face & N.A. \\
3 & rs7617493 & 3080200 & 3p26.2 & 0.4294 & IL5RA (0) & lip and chin and nose & \textcolor{black}{black} \\
3 & rs6779419 & 67740402 & 3p14.1 & 0.4257 & SUCLG2-AS1 (0) & lip and chin & \textcolor{myorange}{orange} \\
3 & rs7643249 & 169813340 & 3q26.2 & 0.4283 & LRRC34 (+0.554kb) & nose & \textcolor{mygreen}{green} \\
3 & rs2080794 & 190271177 & 3q28 & 0.2532 & CLDN1 (-34.52kb) & tophead & \textcolor{myred}{red} \\
3 & rs6771833 & 994203 & 3p26.3 & 0.3326 & CNTN6 (-98.45kb) & left side & \textcolor{mypurple}{purple} \\
7 & rs587741 & 105386461 & 7q22.3 & 0.4911 & SRPK2 (0) & eye and nose tip & \textcolor{mypink}{pink} \\
7 & rs17657924 & 96625589 & 7q21.3 & 0.4416 & LOC100506136 (0) & chin & \textcolor{myyellow}{yellow} \\
7 & rs847375 & 16947921 & 7p21.1 & 0.3071 & AGR3 (+65.93kb) & eye & \textcolor{mycyan}{cyan} \\
10 & rs4980297 & 125275363 & 10q26.13 & 0.3882 & CTBP2 (+114.3kb) & upper and mid face &  \textcolor{mygray}{gray}  \\
2 & rs4675833 & 241947122 & 2q37.3 & 0.2307 & LINC01237 (0) & nose tip and philtrum &  \textcolor{mylotus}{lotus} \\
\bottomrule
\end{tabular}%
}
\end{table}

\section{Discussion}

In this paper, we propose a data-driven trimmed feature screening method based on a tensor ridge regression model via setting thresholds on the tensor coefficients to perform feature screening procedure. The inputs of the TrimTenRidge method are a tensor response along with a high dimensional set of predictors and it outputs all non-zero components of the tensor coefficients. The main contribution of TrimTenRidge can be summarized from theoretical, methodological, and five applicational aspects that are described in the Introduction section. Extensive simulation studies with various difficulty levels demonstrate that the TrimTenRidge approach achieves near 100\% success rates with false negative rates well controlled for, if sample sizes are adequate. Since the proposed approach is different from any of the existing approaches in the literature, we do not compare it with other approaches.

Although adding a LASSO penalty or assuming various sparsity structures to the tensor regression will facilitate variable selection, it is not appropriate for the motivated facial shape-GWAS data for three reasons: 1) In addition to variants with strong effects, there also exist a large number of variants with small but nonzero effects in genomic data \citep{barber2019knockoff, boyle2017expanded}. 2) For a univariate response, the coefficient of one predictor is only a scalar that makes the interpretation much easier. However, for a tensor response, the coefficient of one predictor is also a tensor. In practice, some genes are associated with the entire face but other genes may only be associated with some local small regions. Therefore, it does not make sense to shrink all coefficients of each predictor to zero. 3) Linkage disequilibrium (LD), the nonrandom correlation of alleles at nearby loci, is widespread in genomes, with approximately 70\% to 80\% of genomes showing regions of high LD \citep{carlsen2016exploiting}. The Ridge penalty outperforms other regularization approaches in solving these three problems \citep{carlsen2016exploiting, saleh2019theory}.

We apply the TrimTenRidge approach to the human facial shape-GWAS dataset to detect important genetic factors associated with human facial shape variation. We successfully discovered some new findings in addition to confirming other existing genes that were found to be associated with face related traits. These findings may eventually provide interventions for craniofacial dysmorphology, birth defects, and other clinical and forensic endpoints in the future \citep{liu2012genome, claes2016new, shaffer2016genome, sero2019facial, white2021insights}.

\begin{appendix}
\section*{}\label{appn} 
\subsection*{Proof of Lemma \ref{lemma1}}
\begin{enumerate}
\item Since $\Qbf^T\Qbf = \Id_r$, $\Dbf^2$ contains positive eigenvalues of $\Xbf^T\Xbf$, we know that
\begin{equation*}
\left(\frac{1}{h}\Dbf^2+\Id_r\right)^{-1}
\leqslant
\frac{h/\lambda_1}{1+h/\lambda_1}\Id_r,
\end{equation*}
so
\begin{equation*}
\mathbb{E}[\|bias(\hat{\mathscr{B}})\| |\Xbf]
\leqslant
\|\mathscr{B}\|h\lambda_1^{-1}.
\end{equation*}
Then by conditions (C1) and (C2), we see
\begin{equation*}
\mathbb{E}[(\langle \mathscr{G},bias(\mathscr{B})\rangle)^2|\Xbf]
\leqslant
\mathbb{E}[\|bias(\hat{\mathscr{B}})\|^2 |\Xbf]
=
d_1d_2O(h^2n^{-2(\eta-\tau)})
\end{equation*}
for any $\mathscr{G}\in\mathbb{R}^{p\times d_1\times d_2}$ and
$\| \mathscr{G}\| = 1$. Also, since
\begin{equation*}
\Var[\hat{\mathscr{B}}_{: ij}|\Xbf]
\leqslant
\sigma^2h^{-1}\Id_p,
\end{equation*}
we can then yield the result.

\item For each $i=1,2,\dots d_1$ and $j=1,2,\dots, d_2$, we have
\begin{equation*}
\begin{split}
\mathbb{E}[\|\Xbf\hat{\mathscr{B}}_{: ij}-\Xbf\mathscr{B}_{: ij}\|^2|\Xbf] &=trace [\Xbf\Var[\hat{\mathscr{B}}_{: ij}|\Xbf]\Xbf^T]+\|\Xbf \mathbb{E}[ bias(\hat{\mathscr{B}}_{: ij}|\Xbf)]\|^2 \\
& \leqslant \sigma^2 trace[\Pbf\Pbf^T]+h^2\lambda_1^{-1}\|\hat{\mathscr{B}}_{: ij}\|^2 \\
& = \sigma^2 r + O(h^2n^{-(\eta-2\tau)}).
\end{split}
\end{equation*}
This give us the desired result.
\end{enumerate}

\subsection*{Proof of Theorem \ref{theorem1}}
From proof of Lemma 1 we know for any $k=1,2,\dots,p$,
\begin{equation*}
  \mathbb{E}[ bias(\hat{\mathscr{B}}_{k::})|\Xbf] = O(\|\mathscr{B}_{k::}\|h/\lambda_1) = C_1h/n^{\eta-\tau}.
\end{equation*}
If we use $h=C_2a_n^{-2}(\log\log n) n^\theta$, $u_n = 1+(\log\log n)^{-1}$, then
\begin{equation*}
  \frac{C_1h}{(u_n-1)a_n n^{\eta-\tau}} = \frac{C_3(\log\log n)^2}{n^{\eta-\tau-\theta-3\alpha}}.
\end{equation*}
We can see $\mathbb{E}[ bias(\hat{\mathscr{B}}_{k::})|\Xbf]/[(u_n-1)a_n]\to 0$ uniformly in $k$ when $\alpha <(\eta-\tau-\theta)/3$.

To simplify the writing, denote $\ev_k = (0,0,\dots,0,1,0\dots,0)^T$ as a $p$-dimensional vector where the $k$-th component is 1 and others are 0, and $\av_k^T = \ev_k^T(\Xbf^T\Xbf+h\Id_p)^{-1}\Xbf^T$ as the $k$-th row of $(\Xbf^T\Xbf+h\Id_p)^{-1}\Xbf^T$, then
\begin{equation*}
\begin{split}
  & \mathbb{E} [P(\|\hat{\mathscr{B}}_{k::}-\mathscr{B}_{k::}\|>(u_n-1)a_n |\Xbf)]   \\
   \leqslant & \mathbb{E}[ P(\|[\langle (\Xbf^T\Xbf+h\Id_{p})^{-1}\Xbf^T , \mathscr{E}\rangle_{2,1}]_{k::}\| > (u_n-1)a_n |\Xbf)] \\
   = & \mathbb{E}[P(\|\langle \av_k^T, \mathscr{E}\rangle_{2,1}\|>(u_n-1)a_n | \Xbf)].
\end{split}
\end{equation*}

Since almost surely
\begin{equation*}
  \begin{split}
     |\av_k^T\av_k|^2  & = |\ev_k^T(\Xbf^T\Xbf+h\Id_p)^{-1}\Xbf^T\Xbf(\Xbf^T\Xbf+h\Id_{p})^{-1}\ev_k|^2  \\
       & \leqslant |\ev_k^T(\Xbf^T\Xbf+h\Id_{p})^{-1}\ev_k|^2 \\
       & \leqslant h^{-2}.
  \end{split}
\end{equation*}

Notice $\langle \av_k^T, \mathscr{E}\rangle_{2,1}$ is a combination of sub-Gaussian variables, we have
\begin{equation*}
  \begin{split}
     \mathbb{E} [P(\|\hat{\mathscr{B}}_{k::}-\mathscr{B}_{k::}\|>(u_n-1)a_n |\Xbf)] & \leqslant \mathbb{E}[P(\|\langle \av_k^T, \mathscr{E}\rangle_{2,1}\|>(u_n-1)a_n | \Xbf)] \\
       & \leqslant 2\exp\{-\frac{[(u_n-1)a_n]^2}{2d_1^2d_2^2\sigma^2h^{-2}}\} \\
       & \leqslant 2\exp \{-\frac{c_1n^{2\theta+2\alpha}}{2d_1^2d_2^2\sigma^2}\}.
  \end{split}
\end{equation*}

Hence
\begin{equation*}
  \begin{split}
     P(\mathcal{M}_{\mathscr{B},a_nu_n}\subset \mathcal{M}_{\hat{\mathscr{B}},a_n}) & \geqslant 1-P(\bigcup_{k: \|\mathscr{B}_{k::}\|>u_na_n}\{\|\hat{\mathscr{B}}_{k::}\|\leqslant a_n\}) \\
       & \geqslant 1- P(\bigcup_{k: \|\mathscr{B}_{k::}\|>u_na_n}\{\|\hat{\mathscr{B}}_{k::}-\mathscr{B}_{k::}\|>(u_n-1)a_n\}) \\
       & \geqslant 1-2p\exp \{-\frac{c_1n^{2\theta+2\alpha}}{2d_1^2d_2^2\sigma^2}\}.
  \end{split}
\end{equation*}

Similarly, we have
\begin{equation*}
  \mathbb{E}[P(\|\hat{\mathscr{B}}_{k::}-\mathscr{B}_{k::}\|>(1-u_n^{-1})a_n|\Xbf)]\leqslant 2\exp \{-\frac{c_1n^{2\theta+2\alpha}}{2d_1^2d_2^2\sigma^2}\},
\end{equation*}
and hence
\begin{equation*}
  \begin{split}
     P(\mathcal{M}_{\hat{\mathscr{B}},a_n}\subset \mathcal{M}_{\mathscr{B},a_n/u_n}) & \geqslant P(\bigcap_{k:\|\mathscr{B}_{k::}\|\leqslant a_n/u_n}\{\|\hat{\mathscr{B}}_{k::}\|\leqslant a_n\}) \\
       & \geqslant 1- P(\bigcup_{k:\|\mathscr{B}_{k::}\|\leqslant a_n/u_n}\{\|\hat{\mathscr{B}}_{k::}-\mathscr{B}_{k::}\|>(1-u_n^{-1})a_n\}) \\
       & \geqslant 1-2p\exp \{-\frac{c_1n^{2\theta+2\alpha}}{2d_1^2d_2^2\sigma^2}\}.
  \end{split}
\end{equation*}
So we see
\begin{equation*}
  P(\mathcal{M}_{\mathscr{B},a_nu_n}\subset \mathcal{M}_{\hat{\mathscr{B}},a_n} \subset \mathcal{M}_{\mathscr{B},a_n/u_n}) \geqslant 1 - 4p\exp \{-\frac{c_1n^{2\theta+2\alpha}}{2d_1^2d_2^2\sigma^2}\}.
\end{equation*}
\end{appendix}

\section*{Acknowledgments}
The authors would like to thank the anonymous referees, an Associate
Editor and the Editor for their constructive comments that improved the
quality of this paper.

\section*{Funding}

This work was funded by grants from the National Institute for Dental and Craniofacial Research to Weinberg: U01-DE020078, R01-DE016148, and R01-DE027023.

The second author was supported in part by NSF DMS 1764280 and 1821157.

\section*{Data Availability}
All of the genotypic markers for the 3D Facial Norms dataset are available to the research community through the dbGaP controlled access repository (http://www.ncbi.nlm.nih.gov/gap) at accession number phs000949.v1.p1. The raw source data for the phenotypes (the 3D facial surface models in.obj format) are available through the controlled-access FaceBase Consortium (www.facebase.org). Access to these facial scans requires institutional ethics approval and from the FaceBase data access committee approval.


\section*{Supplementary Material}

\noindent\textbf{Extra simulation results.}

In addition to the results visually demonstrated in Figures 1-6, we provide more quantitative results in the following Tables 4-6. We assess the performance of the TrimTenRidge approach through the following three criteria:
\begin{enumerate}
\item $\mathcal{S}$: the minimum model size that is required to select all the true coefficients. We report the mean (and standard deviation) of $\mathcal{S}$ across 100 replications.
\item $\mathcal{P}_a$: the proportion that all the true coefficients are selected within a pre-given model size across 100 replications. In all simulation studies we set the model size to be 37.
\item $\mathcal{F}$: the false discovery rate, which is defined as the ratio of the number of noise predictors that are being selected versus total number of predictors. We report the mean (and standard deviation) of $\mathcal{F}$ across 100 replications.
\end{enumerate}
\begin{table}[H]
\caption{The mean (and standard deviation) of $\mathcal{S}$, $\mathcal{P}_a$, and $\mathcal{F}$ obtained by the TrimTenRidge approach across 100 replications for Simulation Setting 1.}
\small
\centering
    \begin{tabular}{c|c|c|c}
    \hline
     & $\mathcal{S}$ & $\mathcal{P}_a$ & $\mathcal{F}$ \\
    \hline
    $(n,p)=(200,2000)$ & 8.25 (1.6900) & 1  & 0.2489 (0.1569) \\
    \hline
    $(n,p)=(200,5000)$ & 8.70 (2.2585) & 1  & 0.2361 (0.1764) \\
    \hline
    $(n,p)=(200,10000)$ & 8.47 (1.6481) & 1  & 0.2471 (0.2557) \\
    \hline
    $(n,p)=(500,2000)$ & 7.50 (1.2673) & 1  & 0.0856 (0.0359) \\
    \hline
    $(n,p)=(500,5000)$ & 7.57 (1.1656) & 1  & 0.1094 (0.0839) \\
    \hline
    $(n,p)=(500,10000)$ & 7.92 (1.8073) & 1  & 0.1403 (0.1222) \\
    \hline
    $(n,p)=(1000,2000)$ & 7.12 (0.4330) & 1  & 0.0515 (0.0154) \\
    \hline
    $(n,p)=(1000,5000)$ & 7.4 (0.9744) & 1  & 0.0540 (0.0352) \\
    \hline
    $(n,p)=(1000,10000)$ & 7.51 (1.0298) & 1  & 0.0571 (0.0410)\\
    \hline
    \end{tabular}
\end{table}

As demonstrated in the Table 2, TrimTenRidge approach only needs an average model size around 8 to detect all the seven true coefficients even when sample size is only 200, which is very effective in locating the true positive coefficients. Therefore, without surprise we notice that all the seven true coefficients are 100\% times successfully selected for a pre-determined model size of 37. The false negative rates significantly decreases from 0.24 to 0.05 as sample sizes increase to 1,000.
\begin{table}[H]
    \caption{The mean (and standard deviation) of $\mathcal{S}$, $\mathcal{P}_a$, and $\mathcal{F}$ obtained by the TrimTenRidge approach across 100 replications for Simulation Setting 2.}
    \centering
    \begin{tabular}{c|c|c|c}
         \hline
     & $\mathcal{S}$ & $\mathcal{P}_a$ & $\mathcal{F}$ \\
    \hline
    $(n,p)=(200,50)$ & 10077.13 (22998.01) & 0.68 & 0.0010 (0.0004) \\
    \hline
    $(n,p)=(200,80)$ & 115527 (127740.67) & 0.47 & 0.0005 (0.0002) \\
    \hline
    $(n,p)=(200,100)$ & 260172.70 (250881.12) & 0.35 & 0.0005 (0.0002) \\
    \hline
    $(n,p)=(500,50)$ & 4 (0) & 1 & 0.0017 (0.0005) \\
    \hline
    $(n,p)=(500,80)$ & 4.05 (0.5) & 1 & 0.0006 (0.0003) \\
    \hline
    $(n,p)=(500,100)$ & 10003.96 (70352.36) & 0.98 & 0.0006 (0.0003) \\
    \hline
    $(n,p)=(1000,50)$ & 4 (0) & 1 & 0.0028 (0.0007) \\
    \hline
    $(n,p)=(1000,80)$ & 4 (0) & 1 & 0.0009 (0.0004) \\
    \hline
    $(n,p)=(1000,100)$ & 4 (0) & 1 & 0.0009 (0.0004) \\
    \hline
    \end{tabular}
\end{table}

Simulation setting 2 represents a very challenging scenario because it has a huge pool containing $p^3$ candidate coefficients to choose from. It is easy to miss true coefficients when sample size is only 200. As a result, the average models sizes $\mathcal{S}$ are extremely large and $\mathcal{P}_a$'s are far away from satisfactory. However, as sample size increases to only 1000 or even 500 for some cases, the TrimTenRidge approach achieves perfect performance by having average model size of 4 without any errors and false negative rates less than 0.001.

\begin{table}[H]
    \caption{The mean (and standard deviation) of $\mathcal{S}$, $\mathcal{P}_a$, and $\mathcal{F}$ obtained by the TrimTenRidge approach across 100 replications for Simulation Setting 3.}
    \centering
    \begin{tabular}{c|c|c|c}
         \hline
     & $\mathcal{S}$ & $\mathcal{P}_a$ & $\mathcal{F}$ \\
    \hline
    $(n,p_1)=(200,2000)$ & 17530.94 (91844.74) & 0.5 & 0.2309 (0.0719) \\
    \hline
    $(n,p_1)=(200,5000)$ & 21136.11 (122146.20) & 0.42 & 0.5142 (0.0522) \\
    \hline
    $(n,p_1)=(200,10000)$ & 82431.14 (436438.11) & 0.27 & 0.6731 (0.0519) \\
    \hline
    $(n,p_1)=(500,2000)$ & 15.06 (0.4221) & 1 & 0.1437 (0.1790) \\
    \hline
    $(n,p_1)=(500,5000)$ & 15.55 (5.5) & 0.99 & 0.2696 (0.1778) \\
    \hline
    $(n,p_1)=(500,10000)$ & 15.78 (5.7323) & 0.99 & 0.6891 (0.1493) \\
    \hline
    $(n,p_1)=(1000,2000)$ & 15 (0) & 1 & 0.0012 (0.0019) \\
    \hline
    $(n,p_1)=(1000,5000)$ & 15 (0) & 1 & 0.0005 (0.0006) \\
    \hline
    $(n,p_1)=(1000,10000)$ & 15 (0) & 1 & 0.0003 (0.0003) \\
    \hline
    \end{tabular}
    \label{matrix-response}
\end{table}
As another example with extremely large pool having $p_1\times p_2\times d_1\times d_2$ candidate coefficients to choose from, the results of Simulation setting 3 are similar to those of simulation setting 2. The average models sizes $\mathcal{S}$ are also extremely large and $\mathcal{P}_a$'s are far away from satisfactory. However, as sample size increases to 1000, the TrimTenRidge approach achieves perfect performance by having average model size of 15 without any errors and false negative rates less than 0.0003 for $p_1=10,000$.


\bibliographystyle{plainnat} 
\bibliography{2_2.bib}       

\end{document}